# A Reference Implementation of WECC Composite Load Model in Matlab and GridPACK

Qiuhua Huang, *Member, IEEE,* Renke Huang, *Member, IEEE,* Bruce J. Palmer, Yuan Liu, *Member, IEEE,* Shuangshuang Jin, *Member, IEEE,* Ruisheng Diao, *Senior Member, IEEE,* Yousu Chen, *Senior Member, IEEE,* Yu Zhang, *Member, IEEE*

*Abstract*—The composite load model (CLM) proposed by the Western Electricity Coordinating Council (WECC) is gaining increasing traction in industry, particularly in North America. At the same time, it has been recognized that further improvements in structure, initialization and aggregation methods are needed to enhance model accuracy. However, the lack of an open-source implementation of the WECC CLM has become a roadblock for many researchers for further improvement. To bridge this gap, this paper presents the first open reference implementation of the WECC CLM. Individual load components and the CLM are first developed and tested in Matlab, then translated to the high performance computing (HPC) based, parallel simulation framework - GridPACK™. The main contributions of the paper include: 1) presenting important yet undocumented details of modeling and initializing the CLM, particularly for a parallel simulation framework like GridPACK™; 2) implementation details of the load components such as the single-phase air conditioner motor; 3) implementing the CLM in a modular and extensible manner. The implementation has been tested at both the component as well as system levels and benchmarked against commercial simulation programs, with satisfactory accuracy.

## I. Introduction

DYNAMIC load modeling has been widely recognized to be essential for power system stability studies [1]-[5]. Over the last several decades, significant efforts have been devoted to developing appropriate dynamic load models [1]-[6] and determining the load model components and their parameters [8]-[12]. As a result, considerable progress in dynamic load modeling has been made—the industry load modeling practice has gradually evolved from the simple static load model, to the "interim" three-phase induction motor plus the ZIP load model, and to the most recent composite load model (CLM) [3]-[5].

In North America, development of the WECC CLM has been an important effort in the last decade [3], [5], [6]. It is generally considered as the state-of-the-art [5], mainly due to its capabilities for representing the diversity in the composition and dynamic characteristics of end-use loads and modeling the electrical distance between the end-use devices and the transmission substations [3]. These capabilities are essential to capture the stalling effects of motors which lead to fault induced delayed voltage recovery (FIDVR) [3], [7].

In recent years, significant efforts have been devoted to estimating the WCC CLM parameters. Load model data tools have been developed in [8] and [9] to derive the CLM parameters based on regional end-use survey data and the "rule of association" method. In [10], the analytical similarity of parameter sensitivity was leveraged to efficiently estimate the CLM parameters from measurements. Parameter dependency was analyzed and utilized to improve measurement-based parameter estimation in [11]. A guideline for developing load model composition data was developed by North American Electric Reliability Corporation (NERC) [12].

Compared to efforts in identifying the parameters, efforts on investigating the CLM structure are limited. A new CLM structure was proposed in [13] specifically for industry facilities. The main difference from the WECC CLM is the introduction of a synchronous motor model and a variable frequency drive (VFD) motor model, instead of using a generic performance model for all power electronic devices in the WECC CLM.

While significant progress in model development, parameter estimation, data tools and industry applications in planning studies has been made, it should be noted that more work is needed to improve the current WECC CLM to more accurately represent the aggregated behavior of end-use loads, including but not limited to the following issues [5]:

a) *Single-phase induction motor model:* The existing model is a static performance-based model and is only adequate for single-phase air conditioners (A/Cs) with a reciprocating compressor. This model is also not adequate in terms of modeling the fault point-on-wave [7] and voltage ramping effects [5]. In addition, new A/C motors are mostly equipped with scroll compressors and/or power electronic drive, which means their dynamic characteristics are significantly different. Consequently, a different dynamic model is needed.

b) *Aggregated modeling of protection for motors and power electronic loads:* Given the differences in location, protection types and settings, these devices under the same substation do not stall and/or trip at the same time [14], thus adequate modeling of the diversity in stalling and tripping is needed [5].

Funding for this work was provided by the U.S. Department of Energy's Office of Electricity Delivery and Energy Reliability through its Advanced Grid Modeling Program. Pacific Northwest National Laboratory is located in Richland, WA and is operated by Battelle Memorial Institute under contract DE-AC05-76RLO1830 with the U.S. Department of Energy.

Q. Huang, R. Huang, B. Palmer, Y. Liu, S. Jin, R. Diao, Y. Chen, Y. Zhang are with Pacific Northwest National Laboratory, Richland, WA, 99354 USA (e-mail: {Qiuhua.Huang, Renke.Huang, Bruce.Palmer, Yuan.Liu, Shuangshuang.Jin, Ruisheng. Diao, Yousu.Chen, Yu.Zhang}@pnnl.gov )

*c) Power electronic drive:* Increasing numbers of end-used loads, particularly in industry facilities, are fed by VFD, thus, it is desirable to model electronic drives in more detail [13].

*d) Distributed generation (DG):* Currently a simplified PV model is considered to model DG in WECC CLM, which is operating under a constant power factor mode. Considering the updated grid interconnection requirements, new DG models with advanced control capabilities should be considered.

*e) Modular implementation*: Currently, a fixed structure is used for WECC CLM, as shown in Fig. 1. While a modularized implementation is planned in the next phase development to enhance modeling flexibility [5], it is still not clear that whether the new WECC CLM will be flexible enough to support customized or user-defined load component models, which is necessary for addressing the four issues above.

*f) Consistent implementation in different simulation tools*: While a high-level specification of the WECC CLM was provided in [6], many of the details needed for implementation are missing. Consequently, the initialization method and implementation of some components in major simulation tools are not consistent, and different initialization results were reported in [15].

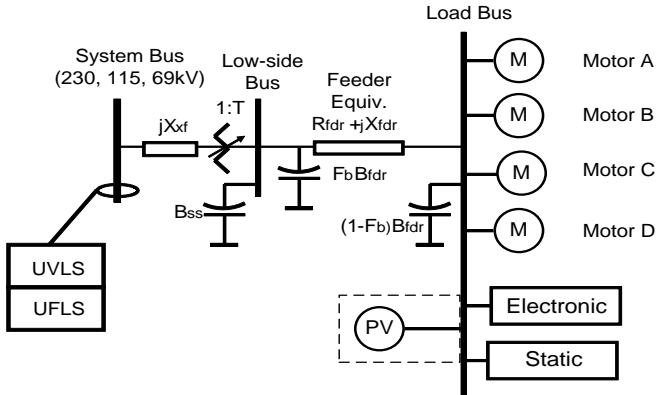

Fig. 1. The structure of the WECC composite load model [6].

The issues of a)-d) mainly pertain to the WECC CLM itself, whereas the last two items are more associated with its implementation in simulation tools. To address issues b) and d), it may require not only improvement on the individual component, but also changes in the model structure. For example, not all the components are aggregated at the end of the equivalent feeder and some other configurations may be more suitable. Although the WECC CLM is available in several commercial simulation tools such as GE PSLF [22] and PowerWorld Simulator [23], their implementation is not transparent to users so that they cannot be modified or extended. This means commercial simulation software is not suitable for researchers and engineers to address the six issues discussed above, therefore, an open implementation of the WECC CLM is needed. In this regard, this paper provides a reference implementation of the WECC CLM. To demonstrate the applicability of the reference implementation, the WECC CLM is also implemented in the open-source HPC-based parallel simulation framework GridPACK[TM] [16],[24],[25] developed by Pacific Northwest National Laboratory (PNNL). As each WECC CLM requires "growing" the system by two buses during initialization for dynamic simulation, a special implementation procedure is developed for GridPACK-like HPC parallel simulation programs. The main contributions of the paper include: 1) helping researchers and engineers better understand the details of the WECC CLM and its implementation; 2) providing a reference implementation that can be easily implemented in any other simulation tools, and most importantly, it can be customized to address the issues discussed above and/or for other related research efforts such as load model calibration.

The remainder of the paper is organized as follows: The overall implementation procedure is discussed in Section II. Development of the dynamic components of the WECC CLM is shown in Section III. Implementation details of WECC CLM are described in Section IV. Test cases and results are presented in Section V. The paper is concluded in Section VI.

## II. THE OVERALL IMPLEMENTATION PROCEDURE

The overall development procedure is shown in Fig. 2. The procedure includes two stages: 1) developing and validating the models in Matlab; 2) implementation and validation in GridPACK[TM]. The main objective of the first stage is to take advantage of the easy-to-debug capability in Matlab to develop simulator-agnostic dynamic load component models and WECC CLM and to employ the play-in concept [17] to validate the models at the component level. At the second stage, all the individual load component models as well as the complete WECC CLM are "translated" and implemented in GridPACK. The last step is to validate the models on test systems.

To facilitate "translating" the model developed in Matlab to a power system dynamic simulator such as GridPACK[TM], all the key methods associated with the load components such as initialization and integration are defined and implemented as if the models were developed in a power system simulator. More details will be discussed in the next section.

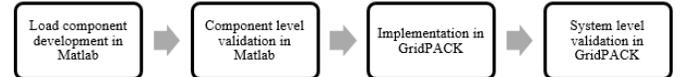

Fig. 2. The overall implementation procedure.

## III. DEVELOPMENT OF DYNAMIC LOAD COMPONENTS

### A. Network Boundary Models of the Load Components

There are four types of load components in the existing WECC CLM, i.e., three-phase induction motor (motor A, B and C in Fig. 1), single-phase induction motor (motor D in Fig. 1), power electronic load and static load. For any dynamic load model implementation, a common and important part of modeling is to determine a proper network boundary model for interfacing the load components with the grid in the network solution step. In this paper, the Norton equivalent model is adopted. The Norton current injection and admittance for different load components are shown in Fig. 3.

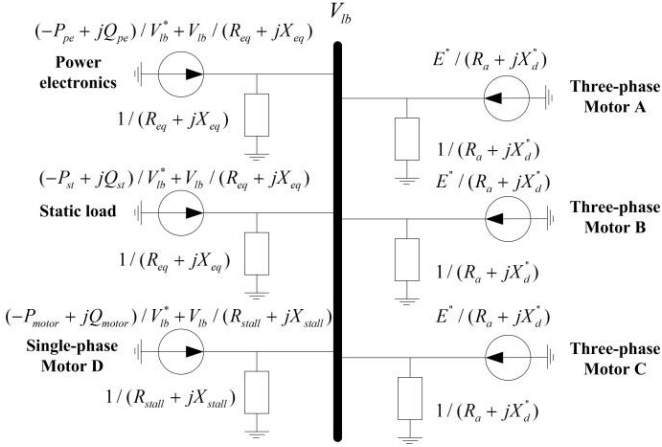

Fig. 3. Network boundary models for dynamic load components.

### B. Key Steps for the Dynamic Load Components

Four key common steps have been identified and must be implemented for the dynamic load components.

1) *Initialization*: This step mainly involves calculating the states variables, if any, the Norton equivalent and the internal variables of the dynamic load components based on the power flow solutions.

2) *Differential equation integration:* If the dynamics of the load components are represented by differential equations, the differential equations are solved for each time step. In this implementation, a predictor-corrector type modified Euler integration method [19] is employed.

3) *Norton current calculation:* The method is called before the network solution and during the iteration of the network solution step to update the equivalent Norton current injections of the load components shown in Fig. 3.

4) *Post-process:* Performance of the load components is significantly influenced by the protections at both the substation and the component levels. Thus, besides the above three methods, a post-processing method should be called to check for tripping conditions and modify models as required before proceeding to the next time step.

### C. Three-Phase Induction Motor

A double-cage induction motor model suitable for electromechanical transient analysis has been developed to represent three-phase motors in the WECC CLM. The block diagram of the five-order motor model [6] is shown in Fig. 4.

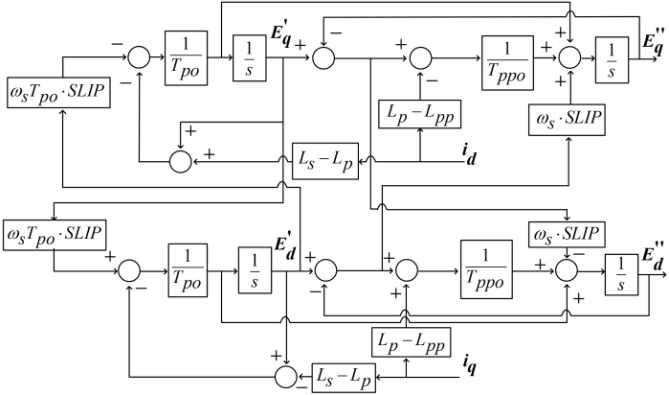

Fig. 4. Block diagram of induction motor model.

In Fig. 4, the four state variables associated with electrical part of motor model are represented by $E_q'$, $E_d'$, $E_q''$ and $E_d''$. The synchronous reactance, transient and subtransient reactances are expressed by $L_s$, $L_p$ and $L_{pp}$. The transient and subtransient rotor time constants are denoted by $T_{po}$ and $T_{ppo}$. *SLIP* represents the per-unit rotor slip. $\omega_s$ is the synchronous frequency. The inputs to Fig. 4 block diagram are terminal currents of the motor, and can be calculated by (1) and (2).

$$i_d = \frac{r_s}{r_s^2 + L_{pp}^2}\left(V_d + E_d''\right) + \frac{L_{pp}}{r_s^2 + L_{pp}^2}\left(V_q + E_q''\right) \quad (1)$$

$$i_q = \frac{r_s}{r_s^2 + L_{pp}^2}\left(V_q + E_q''\right) - \frac{L_{pp}}{r_s^2 + L_{pp}^2}\left(V_d + E_d''\right) \quad (2)$$

The implementation includes two parts: initialization and dynamic process. In the initialization routine, the initial *slip* of the motor (*slip0*) is calculated based on motor *slip-power* curve. The other four state variables $E_q'$, $E_d'$, $E_q''$ and $E_d''$ can be initialized by solving four linear equations, which are created from the corresponding differential equations by assuming the initial derivatives of state variables are zero. Numerical integration steps are executed to update these five state variables. The network boundary conditions, in this case the Norton current injections, are calculated from the state variables and updated in each integration loop. The entire process is illustrated in Fig. 5.

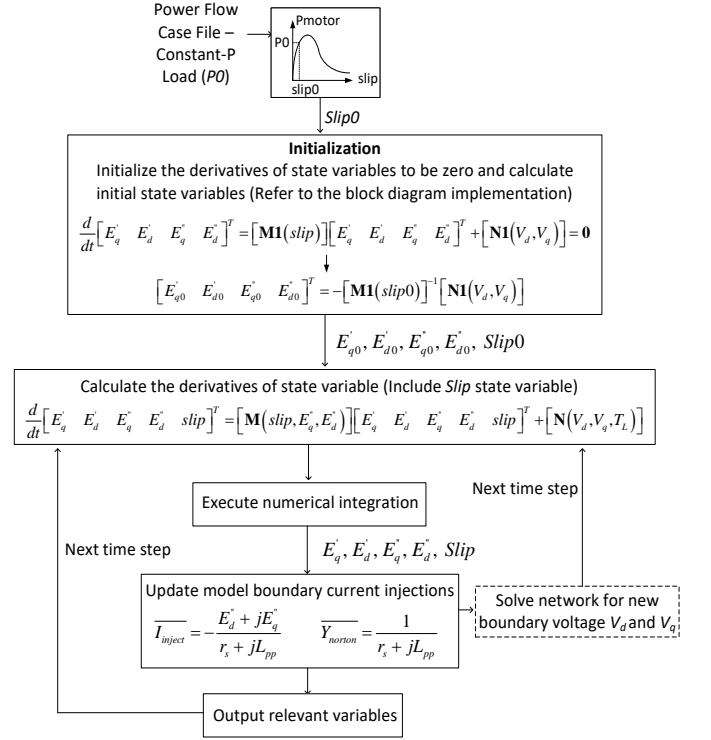

Fig. 5. Entire implementation process of the induction motor model.

### D. Single-Phase Air-Conditioner Compressor Motor

The static performance model used in WECC CLM to represent single-phase residential air-conditioner compressor motors is illustrated in Fig. 6. Algebraic equations for both real and reactive power performance curves can be found in [6]. The performance curve consists of three sections corresponding to three operating states, 1) running state with voltage above breakdown voltage (the black curve in Fig. 6), 2) running state with voltage below breakdown voltage (the red curve in Fig. 6), and 3) stall state (the green curve in Fig.6) [6]. The voltage at the intersection of the last two curves $V_{stallbrk}$ will be computed during initialization. A procedure is provided in [6] to determine $V_{stallbrk}$. However, the tolerance (0.01 pu) used in [6] is found not small enough for a good accuracy. A tolerance of 0.0001 pu is used instead in this paper.

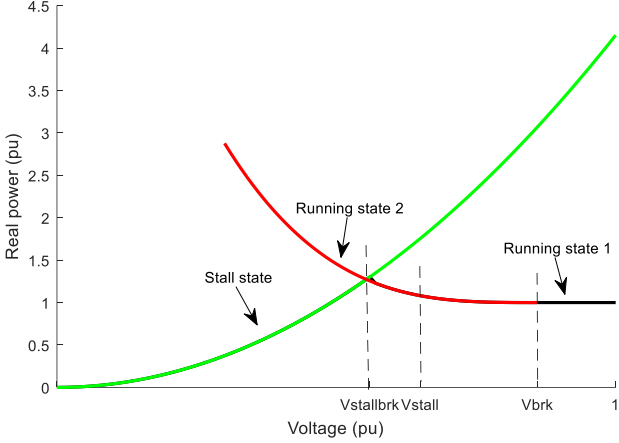

Fig. 6. Power-voltage performance curve of a single-phase residential air-conditioner compressor motor.

Three implementation details should be highlighted:

*1) Restarting and non-restarting A/C motors are separately modeled by two parts:* Considering that the model is a composite of many individual single-phase A/C compressors, and some A/C motors can restart whereas others cannot after stalling within the time frame of simulation, the model is actually internally represented by two parts, i.e., part A for representing "non-restarting" A/C motors and part B for "restarting" A/C motors, as shown in Fig. 7 [6].

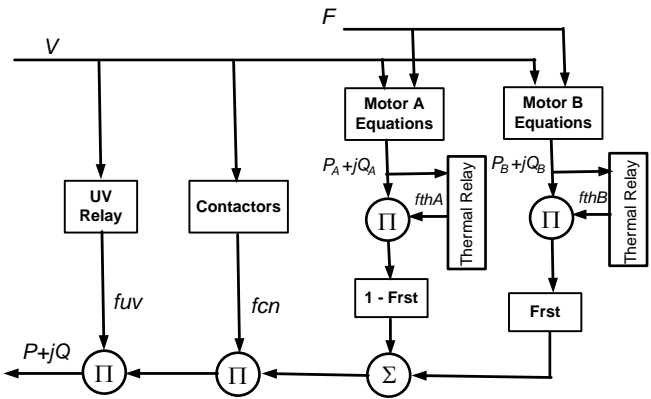

Fig.7. A/C model schematic for illustrating the two-part representation and the protection logics [6].

There are three types of protections equipped in the model, i.e., under-voltage (UV) relay, contractor and internal thermal protection. UV relay and contractor are modeled at the motor model level. Thermal protection is modeled separately for part A and part B, as it is highly related to the motors' restarting capability. Thus, separate variables for the operating status (running or stalling), temperature and thermal protection tripping fraction are used for both part A and B.

*2) Implemented as a variable impedance for running stage 2:* When the A/C motor model is operating in the running state with terminal voltage below breakdown voltage (red curve in Fig. 6), the corresponding power-voltage relationship is by nature unstable, i.e., voltage decrease leads to increased power consumption, which, in turn, further decreases the voltage. The network solution cannot converge due to such load characteristics. Thus, the performance model cannot be strictly enforced in the algebraic network boundary equation solution [18]. To ensure numerical stability, the model is treated like a variable impedance when operating in this state. At each time step, the equivalent impedance is updated based on power consumption and voltage at the last time step, and remains constant for the current time step network solution.

*3) Post-processing step for protection and updating the equivalent impedance:* A post-processing step is proposed for processing the protections, and updating the motor operating status and equivalent impedance if necessary. This step is executed after the network solution is converged and before proceeding to the next time step.

### E. Static Load Model

Equations of the static load component in WECC CLM can be found in [6]. The model is quite standard and is usually used to represent ZIP load models. It is well-known that constant power (CP) and constant current (CI) loads could lead to numerical issues under a low voltage condition [19]. In such a context, constant power and constant current types of loads should be switched to numerically robust representations. In this implementation, the following representation expressed by (3) and (4) is used for CP and CI types of loads, respectively.

$$S_{CP}(V) = \begin{cases} S_{CP}^0 & \text{if } V \geq V_{CP}^{min} \\ \dfrac{S_{CP}^0}{2}(1-\cos(\dfrac{\pi V}{V_{CP}^{min}})) & \text{if } V < V_{CP}^{min} \end{cases} \quad (3)$$

$$S_{CI}(V) = \begin{cases} S_{CI}^0 V & \text{if } V \geq V_{CI}^{max} \\ S_{CI}^0 V \sin(\dfrac{\pi V}{2V_{CI}^{max}}) & \text{if } V < V_{CI}^{max} \end{cases} \quad (4)$$

where $S_{CP}^0$, $S_{CI}^0$ are the initial power (in a complex form) of the CP and CI portions of the static load; $V$ is the load bus voltage magnitude in pu; $V_{CP}^{min}$ and $V_{CI}^{min}$ are the minimum voltage thresholds for maintaining CP and CI models, and their default values are 0.7 pu in this paper. One outstanding advantage of this implementation is that although both representations are piecewise functions, their derivatives are continuous at the piecewise points.

## IV. Development of WECC Composite Load Model

### A. Overview of the Development Approach

Unlike conventional load models, the WECC CLM includes two new buses and two new branches within the model, as shown in Fig. 1. This means that it would be very complicated to fit the WECC CLM into an existing dynamic modeling framework designed for individual components. In this paper, the WECC CLM is treated as a "container" of multiple individual load models. This "container" concept overcomes the issues with the fixed structure of the existing implementation in commercial programs, and supports customization and modeling of additional end-use load models. The implementation of the WECC CLM includes three main steps:

1) After power flow converges, grow the system by adding a low-side bus, a load bus, a distribution transformer and an equivalent feeder for each WECC CLM.
2) For each WECC CLM, run substation-level power flow to initialize the newly added portion of the network and then initialize the dynamic load components connected at the "load bus", which will be discussed in next subsection.
3) Dynamic load components within a WECC CLM are simulated as if they were individual loads except that they all respond to the substation level UV and under-frequency (UF) load shedding protections defined in the WECC CLM.

### B. Initialization of the WECC CLM

First, two new buses and two new branches are added to the system for each composite load. Considering that the original network is needed for some functions, such as power flow, the expansion of this model is triggered during the dynamic simulation initialization stage instead of performing the expansion automatically when parsing the dynamic data file.

Second, the initialization of WECC CLM requires calculating $Tap, V_{ls}, V_{lb}, B_{f1}, B_{f2}, P_{lb}$ and $Q_{lb}$. The associated equations are given in (5).

$$I_{lf} = (P_{lf} - jQ_{lf})/V_{lf}^* \quad (5.a)$$

$$Tap = \sqrt{\frac{V_{lf}^2((V_{min}+V_{max})/2)^2}{(Q_{lb}X_{xfr}-V_{lf}^2)^2+(X_{xfr}P_{lb})^2}} \quad (5.b)$$

$$V_{ls} = (V_{lf} - jX_{xfr}T_{fixhs}^2 I_{lf})T_{fixls}Tap/T_{fixhs} \quad (5.c)$$

$$I_2 = I_1 T_{fixhs}/TapT_{fixls} \quad (5.d)$$

$$I_3 = jB_{ss}V_{ls} \quad (5.e)$$

$$I_4 = I_2 - I_3 \quad (5.f)$$

$$V_{lb} = V_{ls} - (R_{fdr}+jX_{fdr})I_4 \quad (5.g)$$

$$P_{lb} + jQ_{lb} = V_{lb}I_4^* \quad (5.h)$$

The variables in the equations above are either defined in [6] or shown in Fig. 8. In particular, $Tap$ is first calculated based on (5.b), aiming to achieve $V_{ls}=(V_{min}+V_{max})/2$, which is the midpoint of the allowed minimum and maximum voltage levels at the low side bus, then rounded to the closest discreet tap position while enforcing $T_{min}$ and $T_{max}$ constraints [18]. When the calculated $V_{lb}$ is below 0.95 pu, the procedure recommended in [6] is to simply reduce the equivalent feeder impedance to increase $V_{lb}$ to 0.95 pu. While this procedure is easy to implement, it undesirably changes the feeder impedance, which is not only an important physical characteristic of the distribution, but also has high sensitivity on the performance of CLM [20]. In this paper, we propose to always first adjust the distribution transformer tap to increase the voltage profile along the feeder, and only reduce feeder impedance as the last choice. This procedure is closer to actual distribution system operation practice. Finally, the load at the load bus $P_{lb}+jQ_{lb}$ is distributed to the load components and then they are initialized. To simplify the implementation, all the reactive power imbalance caused by dynamic load initialization is compensated by $B_{f2}$ with $B_{f1}$ being set to zero.

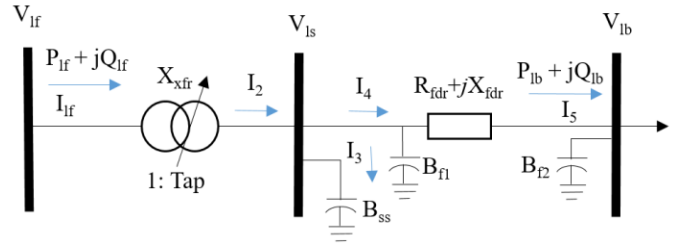

Fig.8. Eqvivalent circuit for initialization of composite load model.

### C. Implementation of the WECC CLM in GridPACK

For a GridPACK-like parallel computing simulator, the main challenge lies in growing the system and properly distributing and indexing the newly added buses and branches, as the power grid network is distributed among the computing nodes. For simplicity, it is desirable that the new buses and branches for each composite load are all located on a single processor and do not have connections to other processors. This can be accomplished by only expanding the load on buses that are owned by a processor and do not represent copies of buses owned by a different processor. All new buses and branches are then solely owned by the processor that owns their parent bus.

The new buses must be given unique bus IDs. To create IDs for the new buses, the highest bus ID for the original buses is determined by first finding the highest ID across processors. New IDs can be created by using integers higher than this maximum value. To do this, it is first necessary to determine how many new buses each processor has. Once each processor knows how many new buses are contributed by all other processors, new IDs can be assigned to the new buses. A similar strategy can also be used to assign new global indices to each bus.

Once bus IDs have been determined, the "from" and "to" buses of the new branches can be assigned and the neighbor lists can be modified to reflect the new geometry. Because the new buses and branches do not have connections to other buses, these operations can all be performed without communi-cating with other processors. At this point the new geometry is complete and can be used for simulation.

## V. TEST CASES AND PERFORMANCE

The dynamic load components developed in Matlab were first tested by using a play-in voltage signal at their terminals. To benchmark the models with that in the commercial simulation software GE PSLF, a two-bus system was created as shown in Fig. 9. The play-in voltage signal is applied at the classical generator model at bus 1. Parameters of the load components and the CLM are provided in the Appendix.

After the composite load model was implemented in Grid-PACK$^{TM}$, it was tested through a system level benchmarking study against another commercial simulation program, PowerWorld Simulator. The test results will be discussed in the following subsection D.

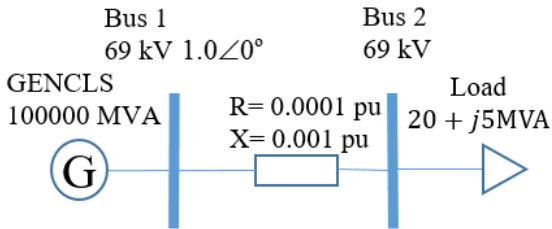

Fig. 9. A two-bus test system.

### A. Three-phase Induction Motor

The responses of the induction motor model developed in this paper and the PSLF MOTORW model are compared considering a voltage sag at the motor terminal. The voltage ramps down from 1.0 pu to 0.5 pu in 0.1 s, then sustains at 0.5 pu for 0.1 s, and finally recovers back to 1.0 pu in 0.1 s. The same voltage sag profile will be used in the following load component tests. The results are shown in Fig. 10. The three-phase induction motor model developed in Matlab is well matched with that in PSLF.

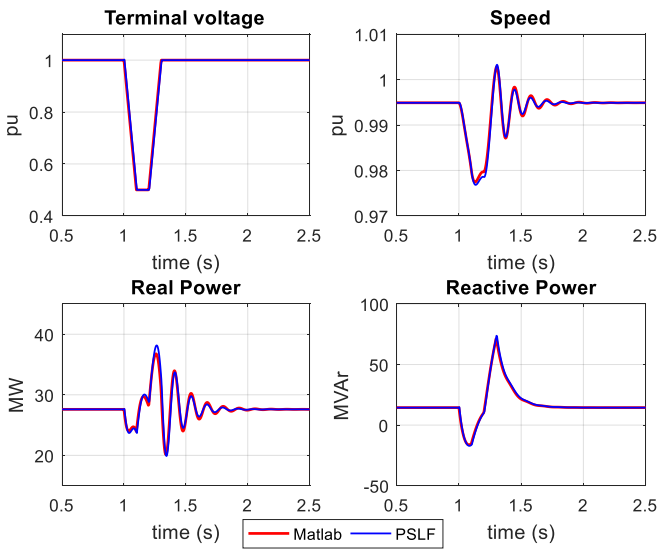

Fig. 10. Comparison of induction motor model developed in this paper and in PSLF.

### B. Single-Phase Air-Conditioner Compressor Motor

The play-in voltage signal is shown in Fig.11(a) and the power consumption of the motor simulated by Matlab and PSLF are compared and shown in Fig.11(b). The voltage sag results in A/C compressor motor stalling. Consequently, the power consumption significantly increases as the terminal voltage recovers. In this test, it is assumed that all A/C motors cannot restart after stalling and the simulation continues till 40 seconds, the motor internal temperature (denoted by TempA) and the fraction of motors (in terms of MVA) not tripped by the thermal protection (denoted by FthA) are shown in Fig. 12. All comparison results are in a good agreement.

### C. Static Load

For the static load model, its implementation under the constant power and constant current modes are specifically tested. The play-in voltage is the same as that shown in Fig.11(a). The voltage thresholds for maintaining constant power and current modes, i.e., $V_{CP}^{min}$ and $V_{CI}^{min}$, are set to 0.70 pu. The comparisons of the actual power of the static load are shown in Fig. 13.

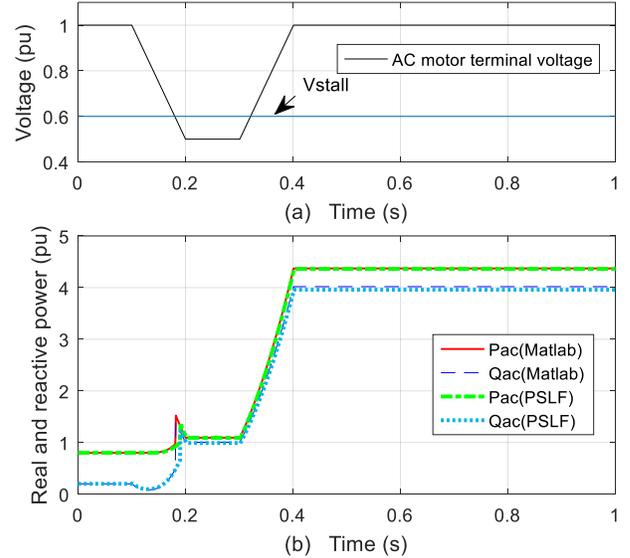

Fig. 11. Comparison of the A/C motor developed in this paper and in PSLF: power consumption.

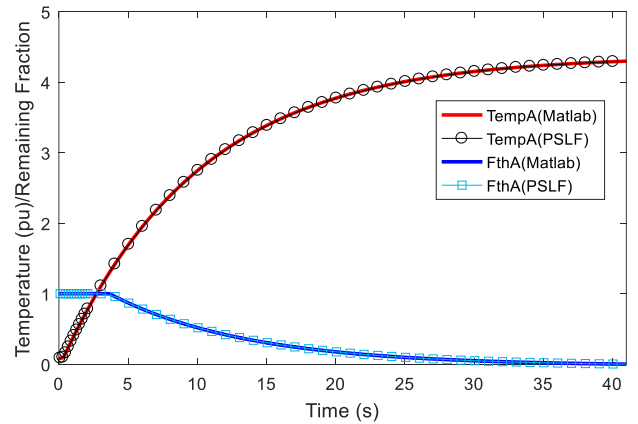

Fig. 12. Comparison of the A/C motor developed in this paper and in PSLF: internal temperature and fraction of motor not tripped by thermal protection.

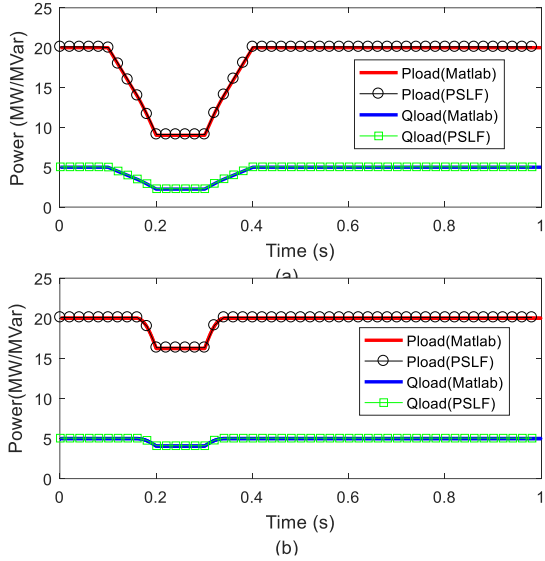

Fig. 13. Comparison results of the static load model: (a) constant current; (b) constant power.

### D. Composite Load Model Implemented in GridPACK$^{TM}$

The composite load model implemented in GridPACK$^{TM}$ has been tested on a modified IEEE 300-bus system [21], with the loads at buses 40, 43, 48, 90, 92, and 105 being modeled by the WECC CLM. Parameters of all CLMs are the same as that in the Appendix. The load composition is 50% three-phase motor A and 30% single-phase A/C motor D and 20% static load. A three-phase fault was applied at bus 90 at 2 seconds and lasted for 0.08 seconds, and the total dynamic simulation was performed for 20 seconds. Fig. 14 to Fig. 17 show the comparisons of the simulation results between GridPACK and PowerWorld, for the low-side bus voltage, the speed of motor A and the real and reactive power of motor D of the composite load at bus 90 as well as the speed of the generator at bus 10007. These figures show a good match between the GridPACK$^{TM}$ and PowerWorld simulation results.

## VI. CONCLUSIONS AND FUTURE WORK

While the WECC CLM is available in several commercial simulation tools such as GE PSLF and PowerWorld Simulator, some important implementation details discussed in this paper are not provided and the model cannot be modified or extended. This paper presents a reference implementation of the WECC CLM in both Matlab and HPC simulation platform GridPACK$^{TM}$. Important implementation details of load components in the WECC CLM and an improved initialization procedure for the WECC CLM are presented. Techniques for addressing the challenges of growing the system and indexing the new buses and branches for modeling the CLM in a distributed network representation in GridPACK$^{TM}$ are also discussed. The load models have been validated by benchmarking against the corresponding models in commercial simulation tools through both component and system level tests.

The WECC CLM represents one of the latest trends in load model development. This work provides a reference and lays a solid foundation for future load modeling research efforts in composite load modeling. Future research directions include: 1) improving the WECC CLM, including but not limited to aggregated protection modeling for each load component, dynamic models for single-phase AC motor and distributed generation resources; 2) load model parameter estimation and calibration. With a validated implementation of the WECC CLM, it is convenient to combine it with the advanced model calibration algorithms developed by PNNL to realize fast and robust estimation of the CLM parameters, particularly for the parameters associated with the aggregated protections.

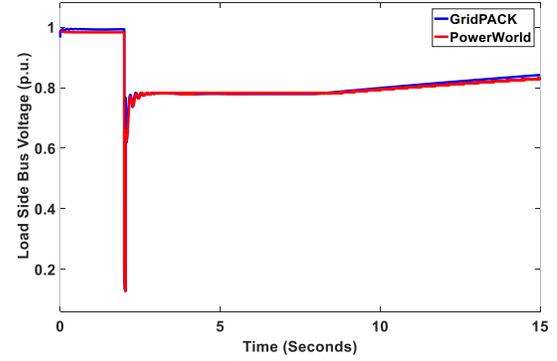

Fig. 14. Comparison of the GridPACK and PowerWorld simulation results for the low-side bus voltage magnitude of the composite load model at bus 90.

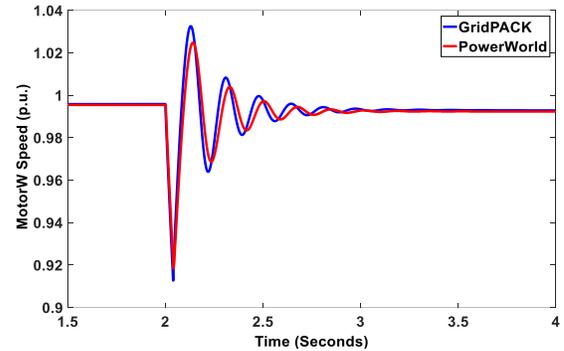

Fig. 15. Comparison of the GridPACK and PowerWorld simulation results for the speed of the motor A of the composite load model at bus 90.

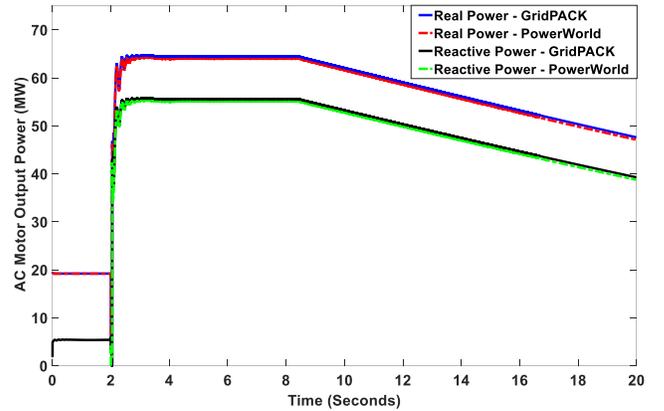

Fig. 16. Real and reactive power of the motor D of the composite load model at bus 90 of the IEEE 300-bus system.

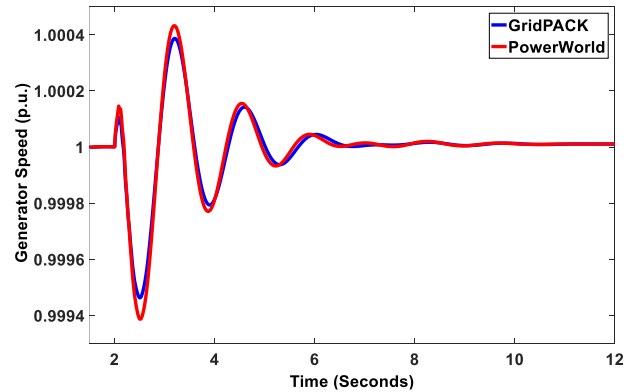

Fig. 17. Comparison of the GridPACK and PowerWorld simulation results for the speed of the generator at bus 10007.

## VII. APPENDIX

The parameters of the composite load model at bus 90 in the PSLF DYD format are provided as follows:

```
cmpldw     90 "90       115.00" 115.0  "1 " : #9 mva= -0.8  "bss" 0.04
"rfdr" 0.0400 "xfdr" 0.0400 "fb" 0.00000 /
 "xxf" 0.0600 "tfixhs" 1.0000 "tfixls" 1.000 "ltc" 0.0000 "tmin" 0.9000
"tmax" 1.1000 "step" 0.006250 /
 "vmin" 1.00 "vmax" 1.0400 "tdel" 30.0000 "ttap" 5.0000 "rcmp" 0.0000
"xcmp" 0.0000 /
 "fma" 0.5 "fmb" 0.00 "fmc" 0.00 "fmd" 0.30 "fel" 0.0000 /
 "pfel" 0.9000 "vd1" 0.8000 "vd2" 0.7000 "frcel" 0.0000  /
 "pfs" 0.90000 "p1e" 2.0000 "p1c" 1.0 "p2e" 1.0000 "p2c" 0.00000 "pfrq" 1.0000 /
 "q1e" 2.0000 "q1c" 1.00000 "q2e" 1.0 "q2c" 0.0000 "qfrq" -1.0000 /
 "mtypa"   3.0 "mtypb"   3.0 "mtypc"   3.0 "mtypd"   1.0 /
 "LFma" 0.800 "Rs" 0.0100 "Ls" 3.1000 "Lp" 0.1779 "Lpp" 0.153900 /
   "Tpo" 1.634 "Tppo" 0.0045 "H" 0.3 "etrq" 0.0000  /
   "vtr1" 0.0 "ttr1" 999 "ftr1" 0.0000 "vrc1" 999.0 "trc1" 999.0 /
   "vtr2" 0.0 "ttr2" 999 "ftr2" 0.0 "vrc2" 999.0 "trc2" 999.0 /
 "LFmb" 0.8000 "Rs" 0.0200 "Ls" 3.6000 "Lp" 0.1800 "Lpp" 0.1800 /
   "Tpo" 1.600 "Tppo" 0.0200 "H" 0.5000 "etrq" 2.0000   /
   "vtr1" 0.80 "ttr1" 2   "ftr1" 1.0000 "vrc1" 1.0000 "trc1" 999.0000 /
   "vtr2" 0.60 "ttr2" 0.16 "ftr2" 1.0000 "vrc2" 999.0000 "trc2" 999.0000 /
 "LFmc" 0.800 "Rs" 0.0200 "Ls" 3.6000 "Lp" 0.1800 "Lpp" 0.1800 /
   "Tpo" 1.600 "Tppo" 0.0200 "H" 0.1000 "etrq" 2.0000   /
   "vtr1" 0.80 "ttr1" 2 "ftr1" 1.0000 "vrc1" 1.0000 "trc1" 999.0000 /
   "vtr2" 0.60 "ttr2" 0.16 "ftr2" 1.0000 "vrc2" 999.0000 "trc2" 999.0000/
 "LFmd" 0.8000 "CompPF" 0.9700  /
   "Vstall" 0.6000 "Rstall" 0.1240 "Xstall" 0.1140 "Tstall" 0.0330 /
   "Frst" 0.0000 "Vrst" 0.9000 "Trst" 999.0   /
   "fuvr" 0.0000 "vtr1" 0.0000 "ttr1" 0.2 "vtr2" 0.0000 "ttr2" 5.0  /
   "Vc1off" 0.45000 "Vc2off" 0.3500 "Vc1on" 0.5000 "Vc2on" 0.4000 /
   "Tth" 10.0000 "Th1t" 1.3 "Th2t" 4.3 "Tv" 0.0500
```

## VIII. ACKNOWLEDGMENT

This work was supported by the U.S. Department of Energy (DOE) through its Advanced Grid Modeling Program. The project team wants to especially thank Mr. Gilbert Bindewald and Mr. Alireza Ghassemian from the U.S. DOE Office of Electricity Delivery and Energy Reliability for their continuing support.